\documentclass[
reprint,
amsmath,amssymb,
aps,
prl,
longbibliography,
floatfix
]{revtex4-2}

\usepackage[english]{babel}

\usepackage{graphicx}
\usepackage[T1]{fontenc}
\usepackage[utf8]{inputenc}
\usepackage[table, dvipsnames]{xcolor}		
\usepackage{verbatim}
\usepackage{dcolumn}
\usepackage{hyperref}
\usepackage[capitalise]{cleveref}
\usepackage[mathlines]{lineno}

\usepackage{svg}
\usepackage{siunitx}						
\usepackage{mathtools}						
\usepackage{physics} 						
\usepackage{braket} 						
\usepackage{MnSymbol}

\renewcommand{\i}{\mathrm{i} }
\newcommand{\e}{\mathrm{e}}
\renewcommand{\d}{\mathrm{d}}
\newcommand{\veps}{\varepsilon}

\newcommand{\eff}{\mathrm{eff}}

\renewcommand{\P}{\mathcal{P}}
\newcommand{\Q}{\mathcal{Q}}

\newcommand{\efield}{\mathcal{E}}
\newcommand{\ddt}{\frac{\mathrm{d}}{\mathrm{d}t}}

\DeclareSIUnit\fs{fs}						
\DeclareSIUnit\eV{eV}						

\begin{document}

\preprint{APS/123-QED}

\title{Quantitative analysis of resonant ionization by smooth laser pulses: \\ Connection between effective Hamiltonian theory and strong-field dressed continua}

\author{Jakob Nicolai Bruhnke
}
\author{Jan Marcus Dahlström$^{\text{\dagger}}$
}

\affiliation{Department of Physics, Lund University, 22100 Lund, Sweden. }

\begin{abstract}
\noindent 
Resonant photoionization in the intense high-frequency regime can exhibit extremely asymmetric Autler–Townes doublets whose origin remains debated. It has been attributed either to interference between perturbative ionization pathways or to a non-perturbative dressing of the continuum. Here we show that these interpretations arise from a common effective Hamiltonian framework, in which dressed-state stabilization is governed by the coherent interplay of resonant and nonresonant pathways. We demonstrate that further simplification, using the strong-field approximation, obscures this mechanism. In contrast, our time-dependent essential-state model, where electrons are rigorously coupled to the continuum, achieves excellent agreement with \textit{ab initio} simulations of the time-dependent Schrödinger equation for helium. 
\end{abstract}

\maketitle

\begin{table}[b!]
\begin{flushleft}
  $^\text{\dagger}$ marcus.dahlstrom@fysik.lu.se
\end{flushleft}
\end{table}

\twocolumngrid
Strong coupling of light and matter is an indispensable phenomenon for  production of novel hybrid systems with highly tunable properties for future applications in chemistry, electronics and quantum science. Quantum electrodynamics of molecules in cavities leads to hybridized controllable chemical reactions with few photons \cite{thomasGroundStateChemicalReactivity2016etal, garcia-vidalManipulatingMatterStrong2021}, while Floquet engineering allows for generation of controllable hybrid quantum states using intense coherent laser fields \cite{wangObservationFloquetBlochStates2013, eckardtColloquiumAtomicQuantum2017, weitenbergTailoringQuantumGases2021, merboldtObservationFloquetStates2025etal, choiObservationFloquetBloch2025}. As intense free-electron laser (FEL) capabilities develop in the high-frequency regime, electrons can be subjected to quantum control on unprecedented scales in energy and time \cite{allariaetal.HighlyCoherentStable2012, emmaetal.FirstLasingOperation2010, nandietal.GenerationEntanglementUsing2024, linkeretal.AttosecondInnershellLasing2025, vismarraetal.DynamicInterferenceChirped2025}. In this way, generation of previously inaccessible systems, such as Floquet-like states in the short-wavelength range will be produced on ultrafast timescales. 
However, the limitations of archetypical control processes, such as Rabi oscillations and adiabatic passage \cite{rabiSpaceQuantizationGyrating1937, morrisTheoryAdiabaticRapid1964, shoreTheoryCoherentAtomic1990}, must be carefully re-examined. 
Rather than being decohered by the environment \cite{zurekDecoherenceEinselectionQuantum2003}, it is now the ultrafast intense light-matter interaction itself that may cause the system to decay -- by breaking it apart via resonant photoionization or autoionization processes following core excitation. Thus, control of the intense field will not only drive population dynamics, but also dynamically reshape the underlying hybrid properties of the system.

While a few landmark experimental achievements have begun to shape the progress of this research field, there persists a knowledge gap in how to interpret the observed phenomena. 
In the extreme ultraviolet (XUV) regime, the achievement of strong coupling was first demonstrated by Nandi {\it et al.} in 2022 at the FERMI FEL for helium atoms \cite{nandietal.ObservationRabiDynamics2022}. Here an observed asymmetry of the avoided crossing, also called the Autler-Townes (AT) doublet, was attributed to photoelectron interference from perturbative ionization processes. Richter \textit{et al.} have recently demonstrated remarkable quantum control using chirped XUV pulses, where selective populations of dressed states was achieved using adiabatic passage techniques \cite{richteretal.StrongfieldQuantumControl2024}. Beyond such population control, strong-field dressing of continuua was reported as the mechanism responsible for asymmetric AT doublets, which implies non-perturbative phenomena beyond time-dependent perturbation theory \cite{saalmannAdiabaticPassageContinuum2018}. These conclusions are in line with simulations of the TDSE in the single active electron (SAE) approximation, which since the 1990s have become a popular tool to study resonant ionization processes \cite{lagattutaAbovethresholdIonizationAtomic1993, girjuNonperturbativeResonantStrong2007, schaferThresholdIonizationHigh1993, mullerBunchingFocusingTunneling1998}. Recent results from Floquet theory indicate however that both continuum and bound states contribute on equal footing to the asymmetry \cite{olofssonCoherentControlIonization2026}, which is in strong tension with the strong-field dressed-continua interpretation of AT asymmetries \cite{saalmannAdiabaticPassageContinuum2018, richteretal.StrongfieldQuantumControl2024}. {Floquet theory provides however only a benchmark for monochromatic fields.} The question arises if it is possible to make quantitative interpretations of such phenomena with{time-dependent essential-states} models.

In this letter, we reconcile these different theoretical interpretations by providing a direct link between strong-field dressing of continua  and interference pathways into the continuum from essential states. {We show} that it is not only the intensity of the field but also the duration of the interaction that determines whether the essential state dynamics are affected by coupling of the continua. Our analytical results and model simulations are in excellent agreement with numerical simulations based on the configuration-interaction singles method (CIS) \cite{foresmanSystematicMolecularOrbital1992, dreuwSingleReferenceInitioMethods2005} and its time-dependent variant (TDCIS) \cite{ greenmanImplementationTimedependentConfigurationinteraction2010, bertolinoThomasReicheKuhnCorrectionTruncated2022}. 
Atomic units are used unless otherwise stated: $e = \hbar =  m = 4\pi\epsilon_0 = 1$.

\begin{figure}[t]
    \centering
    \includegraphics{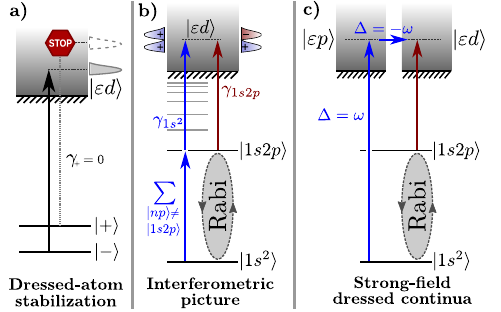}
    \caption{Illustration of stabilization mechanisms. Panel~(a) shows dressed-atom stabilization, where the ionization rate of one dressed state vanishes.  Panel~(b) shows the interferometric picture, where the resonant (red arrow) and non-resonant (blue double arrow) ionization pathways have similar strength and interfere. Panel~(c) shows the strong-field dressing of continua, where the two-photon ionization pathway occurs solely via the continuum, leading to a hybridization of $\ket{\veps p}$ and $\ket{\veps d}$ by plane waves.
    }
    \label{fig:stabcomic}
\end{figure}

\textit{General theoretical approach}---The theoretical foundation for coherent interactions of intense light resonant with single atoms was laid in the 1960s to 1970s \cite{shirleySolutionSchrodingerEquation1965, beersExactSolutionRealistic1975}; we summarize them broadly as \textit{essential states approaches}, where the relevant atomic dynamics is described in a small Hilbert subspace, and the impact of other states---including the continuum---is parametrized perturbatively. This approach has been fruitful for the description of resonance-enhanced multiphoton ionization \cite{holtTimeDependencesTwo1983, dorrTimeEvolutionTwophoton1997, bruhnkeGiantCounterrotatingOscillations2025}, among those we highlight the prediction of dressed-atom stabilization \cite{olofssonPhotoelectronSignatureDressedatom2023} and the interferometric competition of ionization pathways \cite{ishikawaCompetitionResonantNonresonant2012}.

The Hilbert space $\mathcal{H}$ is formally divided into two parts; the essential and nonessential space, $\P$ and $\Q$, respectively, so that $\mathcal{H} = \P\oplus \Q$. We define the usual projection operators $P$, $Q$, expressed in the eigenbasis of the unperturbed Hamiltonian $H_0$, so that $P+Q=1$ \cite{cohen-tannoudjiAtomPhotonInteractionsBasic1998}. For an initial $\P$-space state, $P\ket{\Psi(0)}=\ket{\Psi(0)}$, the exact time-dependent wavefunction can then be evolved using  
\begin{multline}
U(t,0)=\biggl[PU(t,0)P +  \\ -\i \int_0^t \d t' QU^{(\Q)}(t,t')QV(t') PU(t',0)P\biggr]
\label{eq:exactPsi}
\end{multline}
where $V(t)$ is the light-matter interaction, and $U(t,t')$ and $U^{(\Q)}(t,t')$ are the time-evolution operators for exact dynamics in the full Hilbert space and in $\Q$, respectively. 
The former solves $\i \ddt \ket{\Psi(t)} = H(t)\ket{\Psi(t)}$, while the latter solves $\i \ddt Q\ket{\Psi(t)} = QH(t)Q\ket{\Psi(t)}$. The implications of Eq.~\eqref{eq:exactPsi} is that if an {\it exact} solution to the wavefunction is known in the essential states, {\it i.e.} $P\ket{\Psi(t)}$, then the wavefunction in $\Q$ can be obtained through a single coupling $QV(t')P$ followed by propagation solely in $\Q$. 
In essential states approaches, the $\P$-space evolution is described through the use of effective Hamiltonians, $\i \ddt P\ket{\Psi(t)}= H_\eff(t) P \ket{\Psi(t)} $, enabling  
a quantitative interpretation of the underlying physics \cite{friedmannEffectiveTwolevelHamiltonian1978}.

\textit{Dressed-state stabilization}---Beers and Armstrong noted in 1975 that an atom does not always fully ionize in resonant continuous-wave light \cite{beersExactSolutionRealistic1975}. It was recently proposed that a signatures of such dressed-state stabilization is physically realizable in helium atoms using ultrashort XUV-FEL pulses  \cite{olofssonPhotoelectronSignatureDressedatom2023, olofssonCoherentControlIonization2026}. 
We refer to the ground state (GS) as $\ket{a}$ and the excited state (ES) with $\ket{b}$; they have amplitudes $a(t)$ and $b(t)$ respectively. The essential states are coupled by the frequency $\omega_0$ with the corresponding detuning $\delta=E_b-E_a-\omega_0$ with a field amplitude $\efield=-\omega_0 A$.  
The general effective Hamiltonian {for a monochromatic field}, within the pole approximation, reads
\begin{equation} \label{eq:fullHeff}
    H_\eff = \begin{pmatrix}
        S_a - \frac{i}{2} \gamma_a & (\Omega + \i \beta)/2 \\
        (\Omega + \i \beta)/2 & \delta + S_b - \frac{i}{2} \gamma_b
    \end{pmatrix}
\end{equation}
with Stark shifts $S_a$, $S_b$, atomic decay rates $\gamma_a$ and $\gamma_b$, and real and imaginary Rabi couplings $\Omega$ and $\beta$ \cite{beersExactSolutionRealistic1975, holtTimeDependencesTwo1983}.  
%
While the atomic decay rates describe depletion of the essential states, the role of $\beta$ is more subtle. 
The corresponding decay rates of dressed states, $\ket{\pm}$ with eigenvalues $\lambda_\pm$, are determined by the imaginary parts of the eigenvalues. They can be written as $\Im(\lambda_\pm) = -\frac{\gamma_a+\gamma_b}{4} \pm \frac{1}{2}\Im(W)$, where the first term is common to both dressed states, while the second term differentiates the rates. 
Explicitly, the latter term is the imaginary part of the complex generalized Rabi frequency,   
\begin{equation}
\Im(W) = \Im \sqrt{[S_a - S_b - \delta - \i (\gamma_a - \gamma_b)/2]^2 + (\Omega + \i \beta)^2},
\label{eq:ImW}
\end{equation}
which clearly depends on the $\beta$-parameter. 

\textit{Interferometric picture}---
According to perturbation theory, the $\beta$-parameter is part of the third-order coupling of the two essential states via all intermediate states. 
In early works it was assumed that $\beta=\pm\sqrt{\gamma_a\gamma_b}$, but such construction is not valid in the general case with multiple continua \cite{olofssonPhotoelectronSignatureDressedatom2023}. 
By complex coupling of the essential states, the $\beta$-parameter incorporates back-action from the continua, causing different decay rates of the two dressed states, $\Im(\lambda_+) \neq \Im(\lambda_-)$. 
Under intense photoionization, such decay has been connected to the asymmetry in the AT doublet of photoelectrons  \cite{girjuNonperturbativeResonantStrong2007,olofssonPhotoelectronSignatureDressedatom2023, csehiExactAnalyticCharacterization2026}, while for perturbative photoionization the asymmetry has been linked to interference between pathways from the two essential states 
\cite{nandietal.ObservationRabiDynamics2022,zhangEffectNonresonantStates2022}. 
If only one continuum is accessible, destructive interference of two pathways into the continuum with equal amplitudes, $\gamma_a\simeq\gamma_b$, provides a compelling physical interpretation for why either $\Im(\lambda_+)$ or $\Im(\lambda_-)$ will vanish at resonance $S_a-S_b\simeq \delta$ in the effective Hamiltonian description, see Eq.\eqref{eq:ImW} and Ref.~\cite{olofssonPhotoelectronSignatureDressedatom2023}. 

\begin{table*}
    \centering
    \caption{Intensity (in $\mathrm{W/cm^2}$) for dressed-atom stabilization in hydrogen and helium, based on exact Floquet calculations, effective Hamiltonian (eff. Ham.) calculations within the pole approximation with the interferometric condition $\gamma_a = \gamma_b$, and the plane-wave condition of Eq.~\eqref{eq:adiabaticpassagecondition_saalmann}. The transition is $1s \leftrightarrow 2p \ (m=1)$ in hydrogen (H) and $1s^2 \leftrightarrow 1s2p \ (m=1)$ in helium (He), the latter done with an SAE potential \cite{richteretal.StrongfieldQuantumControl2024} and CIS. We compare length gauge (LG) and velocity gauge (VG).}
    \label{tab:conditiontable}
    \begin{tabular}{lcccccc}
       & H (LG) & H (VG) & He (SAE, LG) & He (SAE, VG) & He (CIS, LG) & He (CIS, VG) \\ \hline
       Floquet & $3.2\times 10^{14}$ & $3.2\times 10^{14}$ & $6.9 \times 10^{14}$ & $6.9 \times 10^{14}$ & $1.1\times 10^{14}$ & $1.9\times 10^{14}$ \\
      Eff. Ham. & $3.3 \times 10^{14}$ & $2.4 \times 10^{14}$ & $4.2  \times 10^{14}$ & $3.9\times 10^{14}$ & $8.4\times 10^{13}$ & $1.2\times 10^{14}$ \\
      Plane-wave & --- & $9.9 \times 10^{15}$ & --- & $1.0\times 10^{16}$ & --- & $6.8 \times 10^{15}$ \\
    \end{tabular}
\end{table*}

\textit{Strong-field dressed continua}---Recently, it was reported that the physical mechanism for asymmetries of the AT doublet in intense XUV light requires strong-field effects beyond perturbation theory \cite{richteretal.StrongfieldQuantumControl2024}. The theory of strong-field dressing of two continua was proposed by Saalmann {\it et al.} in 2018, where coherent control of photoionization was predicted by usage of chirped pulses \cite{saalmannAdiabaticPassageContinuum2018}.
Employing the strong-field approximation (SFA) with plane waves $\ket{\mathbf{k}}$, which implies that the Coulomb interaction between the photoelectron and the ion can be neglected, leads to a simple condition for the maximum AT asymmetry for unchirped pulses 
\begin{equation} \label{eq:adiabaticpassagecondition_saalmann}
    \frac{A k}{2\omega}|\braket{k|a}_r| = {\cal C}|\braket{k|b}_r|,
\end{equation}
where $\braket{k|\,\cdot\,}_r$ is the radial part of the essential state momentum wavefunction, and $k = \sqrt{2(2\omega_0 - I_P)}$ the momentum at resonance, with the binding energy $I_P$. We find that this expression is valid for both linear and circular polarization with ${\cal C}= \sqrt{3}$; and that it is consistent with the one-dimension case derived in  Ref.~\cite{saalmannAdiabaticPassageContinuum2018} for $\mathcal{C} = 1$. 

\textit{Reconciliation of physical interpretations}---We illustrate the three physical interpretations of AT asymmetry in Fig.~\ref{fig:stabcomic}. In panel~(a) the dressed-state picture implies distinguishable pathways for the two final energies with one ``stopped'' by its nullified decay rate. In panel (b) the interferometric viewpoint implies two pathways leading to the same final energy, but with destructive and constructive phases, respectively. In panel~(c) we illustrate the strong-field dressed-continuum viewpoint with hybridization of  angular momenta in the continua.   

We will now show that the strong-field dressed-continuum condition in Eq.~\eqref{eq:adiabaticpassagecondition_saalmann} can be directly derived from the effective Hamiltonian. We consider linear polarization in the $z$-direction, but our conclusions are general. The ionization rates in Eq.~\eqref{eq:fullHeff} for the essential states in velocity gauge read
\begin{align}
    \gamma_a &= -\frac{A^4}{8} \Im \left(\sum_{l,m,n}\!\! \frac{\braket{a | p_z | l}\braket{l | p_z | m}\braket{m | p_z |n}\braket{n | p_z | a}}{(E_l - \omega_0)(E_m - 2\omega_0)(E_n - \omega_0)} \right), \\ 
    \gamma_b &= -\frac{A^2}{2} \Im \left( \sum_n \frac{\braket{b | p_z | n}\braket{n | p_z | b}}{E_n - 2\omega_0}\right),
\end{align}
where the sums over $l,m,n$ include bound and continuum states. To connect with the SFA, we now substitute all intermediate states as plane waves. After some algebra we obtain
\begin{align}
    \gamma_a^\mathrm{SFA} = \frac{\pi A^4 k^5}{24} |\braket{a | k}_r|^2 I_\Omega 
    \\  
    \gamma_b^\mathrm{SFA} =  \frac{\pi A^2k^3}{2} |\braket{b | k}_r|^2I_\Omega,
\end{align}
where $I_\Omega$ are identical angular integrals, see the Supplemental Material (SM) for a detailed derivation  \cite{SeeSupplementalMaterialStabilization}. Using the interference criterion, $\gamma_a = \gamma_b$, then yields Eq.~\eqref{eq:adiabaticpassagecondition_saalmann}, which proves that the condition for maximum AT asymmetry by strong-field dressing of continua is based on approximate  elements of the effective Hamiltonian. Further, the use of SFA results in a single hybridized continuum, because $\beta^\textrm{SFA}=-\sqrt{\gamma_a^\mathrm{SFA}\gamma_b^\mathrm{SFA}}$, which is incorrect for linear polarization and
leads to overestimation of the stabilization phenomenon. Thus, we have found that diagonalization of the effective Hamiltonian, with the true $\beta$-parameter included, achieves strong-field dressing of the continua beyond perturbation theory with the result being a quantitative re-balancing of the decay rates of the dressed states. 

In Table~\ref{tab:conditiontable}, the stabilization intensity in hydrogen ($1s \leftrightarrow 2p$) and helium ($1s^2 \leftrightarrow 1s2p$) in circular polarization are presented. We compare Floquet theory, effective Hamiltonian theory (with the condition $\gamma_a = \gamma_b$), and the plane-wave approximation, Eq.~\eqref{eq:adiabaticpassagecondition_saalmann}. 
Strong-field effects refer to cases with a large ponderomotive energy compared to the binding energy of the atom, $U_\mathrm{P} = \efield^2/4\omega^2\gg I_\mathrm{P}$. 
Thus a small Keldysh parameter $\gamma_\mathrm{Keldysh}=\sqrt{I_\mathrm{P}/2U_\mathrm{P}}$ is an indicator for strong-field effects, which for $I_\mathrm{P}\approx\omega_0$ would also imply a breakdown of perturbation theory \cite{reissTheoreticalMethodsQuantum1992}. 
Notably, Ref.~\cite{nandietal.ObservationRabiDynamics2022, richteretal.StrongfieldQuantumControl2024} were performed in parameter regimes with $\gamma_\mathrm{Keldysh} > 10$, which draws into question the validity of the SFA. Table~\ref{tab:conditiontable} corroborates this doubt, with the SFA consistently overestimating the stabilization intensity by at least one order of magnitude.    

Further, it should be noted that time-dependent essential state models {\it without} the $\beta$-parameter also can be used to produce asymmetric AT doublets, {\it c.f.} Ref.~\cite{nandietal.ObservationRabiDynamics2022,zhangEffectNonresonantStates2022}. This suggests that strong-field dressing of the continua is not required for the observed asymmetries of AT doublets. Thus we now resolve this paradox by noting that perturbative photoionization would not alter the essential state dynamics and, therefore,  the view of photoelectron interference would remain valid. Longer interactions, $\gamma_{a/b}\tau\gtrsim 1$, 
would imply a conspicuous role of strong-field dressed continua, or, more precisely, the effect of the $\beta$-parameter on the essential states, and consequently, by Eq.~\eqref{eq:exactPsi}, the photoelectron distributions.

A more subtle point is the question of gauge. The Floquet results, which are exact for 
SAE models, generally align well with the effective Hamiltonian results. The latter are gauge dependent due to truncation of the perturbative expansion; with the length gauge exhibiting best agreement with Floquet theory. 
The CIS approach involves a truncation of the many-body basis, which naturally leads to gauge-dependent Floquet results. Overall, the length gauge is known to be more meaningful when studying populations of bound states \cite{kobeGaugeInvariantFormulation1978} and dressed-continuum states with TDCIS \cite{bertolinoThomasReicheKuhnCorrectionTruncated2022}.

{\it Comparison of numerical and model results---}We now return to opening question: Can {time-dependent} essential state models be used to quantitatively interpret photoionization dynamics from atoms in strong coupling? In particular, we need to demonstrate that the effective Hamiltonian approach is able to quantitatively model the quantum control of photoelectrons from helium $1s^2 \leftrightarrow 1s2p$ ($m=1$) in an intense circularly polarized Gaussian pulse \cite{richteretal.StrongfieldQuantumControl2024}. 
The {
amplitudes $a(t)$ and $b(t)$ are obtained by propagating with} 
a slowly-varying effective two-level Hamiltonian $H_\eff(t)$ in the length gauge. To obtain $H_\eff(t)$, time-independent effective Hamiltonians were calculated from Rayleigh-Schrödinger perturbation theory for a range of frequencies $\omega$ and field strengths $\efield$, yielding $H_\eff(\omega, \efield)$. Atomic parameters were obtained from the CIS method with exterior complex-scaling \cite{simonDefinitionMolecularResonance1979}. Going beyond earlier works \cite{nandietal.ObservationRabiDynamics2022,
zhangEffectNonresonantStates2022,
olofssonPhotoelectronSignatureDressedatom2023,  csehiExactAnalyticCharacterization2026}, we have verified convergence by constructing the effective Hamiltonian in $8$th order without applying the pole approximation {\cite{bruhnkeReconciliationEffectiveHamiltonians2026, MoreDetails}}. We employ a Gaussian envelope $\efield(t) = \efield_0 \exp\{-2\ln(2)[t/\tau]^2\}$ and a linear chirp $\omega(t) = \omega_0 + 2\alpha t$ centered around $\omega_0$, where $\alpha$ is the linear chirp rate.

\begin{figure}[t]
    \centering
    \includegraphics{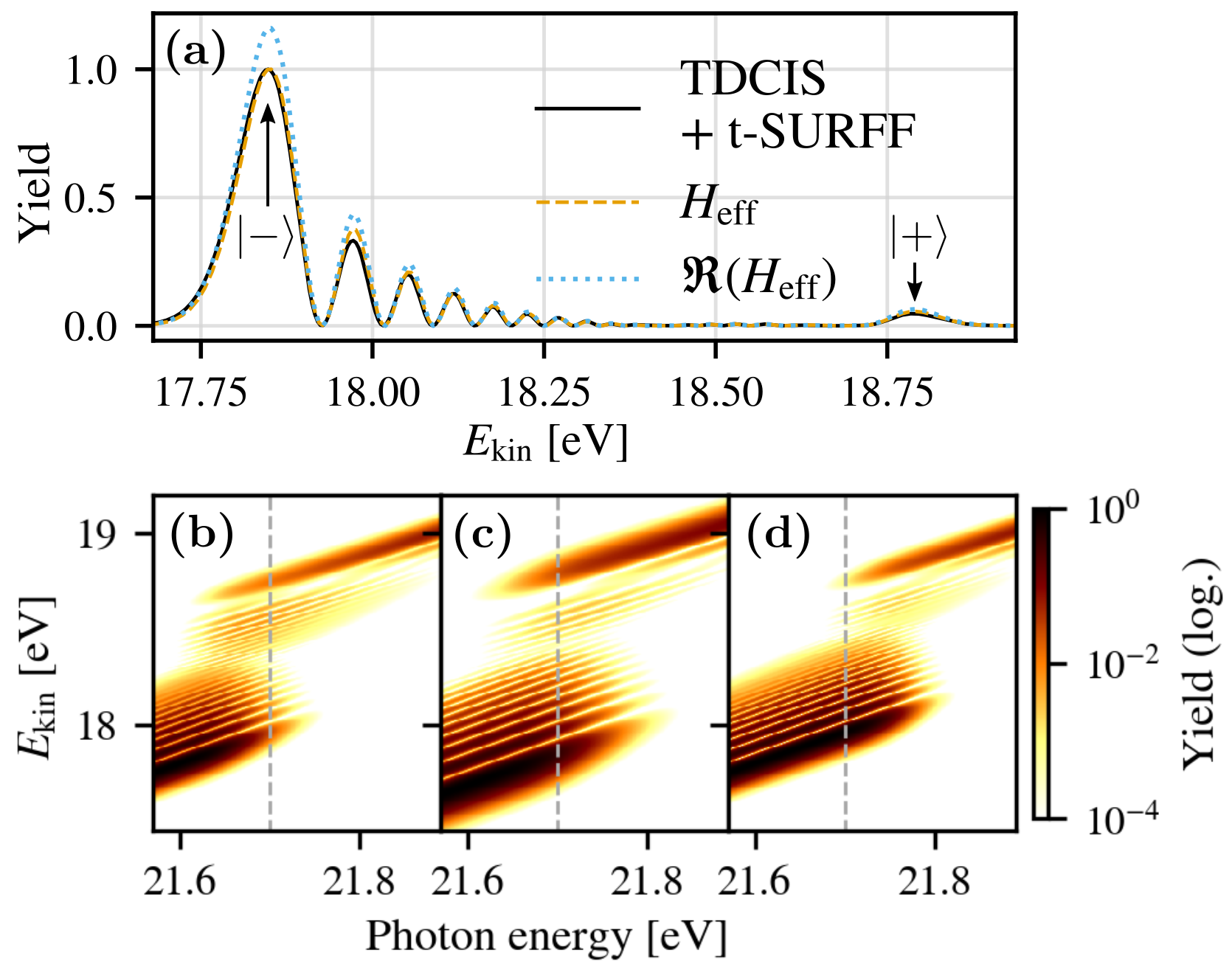}
    \caption{PES obtained for the helium $1s^2 \leftrightarrow 1s2p$ transition with a Gaussian envelope ($I_0 = \SI{3e14}{\watt\per\square\cm}$ and $\tau = \SI{50}{\fs}$, both transform-limited) and three different chirp values, (a) and (c) unchirped, (b) $\mathrm{GDD} = \SI{-1000}{\fs\squared}$, (d) $\mathrm{GDD} = \SI{1000}{\fs\squared}$. Panel (a) compares the PES at the field-free resonance, $\omega_0 = \SI{21.69}{\eV}$ from CIS marked by the vertical lines in the lower row, at different levels of theory: a t-SURFF spectrum in solid black, and the model results in dashed yellow and dotted blue. The model results differ in how the essential state amplitudes are obtained; once through propagation with the full effective Hamiltonian $H_\eff(t)$ and once through propagation with a purely real effective Hamiltonian $\Re[H_\eff(t)]$. The two essential states PES are normalized with the same factor. Panels (b)-(d) show the PES from the essential-states model with decay for varying photon energy $\omega_0$.}
    \label{fig:spectra}
\end{figure}

In Fig.~\ref{fig:spectra} we show photoelectron spectra obtained from the essential states dynamics of $H_\eff(t) \equiv H_\eff(\omega(t), \efield(t))$ by coupling from $\mathcal{P}$ to $\mathcal{Q}$ using Eq.~\eqref{eq:exactPsi}. We choose a transform-limited peak intensity of $I_0 = \SI{3e14}{\watt\per\square\cm}$ and a transform-limited full-width at half-maximum of $\tau = \SI{50}{\fs}$, similar to Ref.~\cite{richteretal.StrongfieldQuantumControl2024}. The asymptotic photoelectron amplitudes are found analytically (to second-order),
\begin{align} \label{eq:PES1}
c_\veps^{(1)} (t) &= -\i \! \int_{-\infty}^t \hspace{-5pt} \d t' \e^{-\i \Delta_\veps (t-t')} \e^{2\i \alpha (t^2 - t'^2)} V_{\veps b}^{(1)}(t') b(t') \\
c_\veps^{(2)} (t) &= -\i \! \!\int_{-\infty}^t \hspace{-5pt} \d t' \e^{-\i \Delta_\veps (t-t')} \e^{2\i \alpha (t^2 - t'^2)} V_{\veps a}^{(2)}(t') a(t'), \label{eq:PES2}
\end{align}
where $\Delta_\veps = \veps + I_P - 2\omega_0$ is the detuning of the photoelectron. Further, $V_{\veps b}^{(1)}$ is the single-photon matrix element and through adiabatic elimination, the two-photon matrix element $V_{\veps a}^{(2)} = \sum_{q\in \Q} V_{\veps q}^{(1)} V_{qa}^{(1)} / (\veps + I_P - E_q - \omega_0)$ was obtained with the atomic energy $E_q$, see SM for details  \cite{SeeSupplementalMaterialStabilization}. The Stark shift of the continuum was incorporated as $\exp[ -\i \int_{t'}^t \d\tau U_P(\tau) ]$ \cite{bagheryEssentialConditionsDynamic2017}.  
The photoelectron spectrum (PES) is then obtained as $|\sum_i c_\veps^{(i)}(\infty)|^2$.

In Fig.~\ref{fig:spectra}(a), we show the PES for an unchirped Gaussian pulse at resonance. We compare our model to {\it ab initio} numerical simulations using TDCIS with the time-dependent surface flux (t-SURFF) method \cite{taoPhotoelectronMomentumSpectra2012, bertolinoPropensityRulesInterference2020}. For the model, we propagate the essential states both with the complex $H_\eff(t)$ and with its real part, $\Re[H_\eff(t)]$. 
The spectrum obtained from $H_\eff(t)$ is in excellent agreement with the
{\it ab initio}
spectrum; showing a strong left AT peak from $\ket{-}$, with pronounced interference fringes that depend on the envelope \cite{rogusResonantIonisationSmooth1986, simonovicManifestationsRabiDynamics2023}, and a faint right AT peak, corresponding to the stabilized $\ket{+}$ state. Tuning the peak intensity closer to the correct stabilization intensity in Table \ref{tab:conditiontable} makes the latter peak weaker, as shown in the SM \cite{SeeSupplementalMaterialStabilization}. The spectrum from $\Re[H_\eff(t)]$ also captures all spectral features, but with slightly higher peak values. Physically, this is expected since decay of the essential states was artificially removed. Similar level of agreement is found for chirped pulses. We conclude that the physics in the investigated parameter regime can be described by an effective Rabi model without decay, {\it i.e.} perturbative photoionization. 

{Accounting for the decay dynamics becomes vital when depletion is substantial, for instance through longer or more intense pulses. As an example, using a longer unchirped pulse with $\tau = \SI{250}{\fs}$, with parameters otherwise left unchanged from Fig.~\ref{fig:spectra}(a), the ratio $R$ between the $\ket{-}$ and $\ket{+}$ peak is predicted as $R = 2.2$ when the inner dynamics is generated by $H_\eff(t)$ and $R = 4.4$ with $\Re[H_\eff(t)]$. Propagating with a complex $H_\eff^{(\beta=0)}(t)$ where $\beta = 0$ is set artificially, we obtain $R = 4.0$, which illustrates the significant role of $\beta$ for the outgoing photoelectrons. It is unphysical to selectively omit the $\beta$ parameter from the effective Hamiltonian, which for (1+1)-resonant photoionization appears in third order of perturbation theory, whereas $\gamma_b$ and $\gamma_a$ appear in second and fourth order respectively \cite{holtTimeDependencesTwo1983}. Such omission is only acceptable when the decay as a whole is negligible, as is the case in Fig.~\ref{fig:spectra}(a). Due to its role in balancing the decay rates of the dressed states, we expect the $\beta$ parameter to play an even more important role when the populations of the dressed states are uneven, for instance when the pulse is detuned or chirped.}

In Fig.~\ref{fig:spectra}(b-d), we show the PES for varying $\omega_0$ with 
$H_\eff(t)$ for chirped Gaussian pulses: 
(b) $\mathrm{GDD} = \SI{-1000}{\fs\squared}$, (c) $\mathrm{GDD} = \SI{0}{\fs\squared}$, and (d) $\mathrm{GDD} = \SI{1000}{\fs\squared}$. Compared to the transform-limited case, the peak intensity is reduced for the chirped cases by a factor of $\sim 0.65$; similarly, the pulses length increases for the two chirped cases by a factor of $\sim 1.5$. 
The shape of the AT doublet depends on two main factors: {\it i)} The populations of the dressed states, and {\it ii)} the decay rates 
of the dressed states. While the chirp has little impact on the decay rates, it significantly affects the population dynamics by selecting a dressed state \cite{saalmannAdiabaticPassageContinuum2018}. 
In Fig.~\ref{fig:spectra}~(b) we show that a large negative chirp can make the $\ket{+}$ AT peak prominent at resonance, despite $\gamma_+ \ll \gamma_-$, because the $\ket{-}$ population is significantly suppressed.

{\it Conclusion---}We have {reconciled} two conflicting views on dressed-state stabilization of atoms: interference between ionization pathways and strong-field dressing of {continua}. Within an effective Hamiltonian framework, we derive the strong-field dressed-continua perspective from the interferometric condition using the SFA. Moreover, we show that the SFA leads to an inaccurate hybridization of continua with exaggerated stabilization intensities beyond present experimental capabilities. In contrast, our time-dependent effective Hamiltonian model, where the essential-state dynamics is rigorously separated from the ionization process, predicts photoelectron spectra that are in excellent agreement with \textit{ab initio} simulations and stabilization at experimentally attainable intensities. 

\textit{Acknowledgements}---We acknowledge discussions with Edvin Olofsson, Yijie Liao, and Mattias Bertolino. 
JMD acknowledges support from the Knut and Alice Wallenberg Foundation: 2024.0212 and the Swedish Research Council: 2024-04247.

\bibliography{MyLibrary}

\end{document}